\documentclass[letter,11pt]{article}
\usepackage{jheppub}

\title{\boldmath Determination of the top quark mass circa 2013:\\ methods, subtleties, perspectives}

\author[a,b]{Aurelio Juste}
\author[c,d]{Sonny Mantry}
\author[e]{Alexander Mitov}
\author[f,g]{Alexander Penin}
\author[e]{Peter Skands}
\author[h]{Erich Varnes}
\author[i]{Marcel Vos}
\author[j]{Stephen Wimpenny}

\affiliation[a]{Instituci\'{o} Catalana de Recerca i Estudis Avan\c{c}ats (ICREA), 08010 Barcelona, Spain}
\affiliation[b]{Institut de F\'{i}sica d'Altes Energies, Universitat Auton\`{o}ma de Barcelona, 08193 Bellaterra, Spain}
\affiliation[c]{High Energy Division, Argonne National Laboratory, Argonne, IL 60439, USA}
\affiliation[d]{Department of Physics and Astronomy, Northwestern University, Evanston, IL 60208, USA}
\affiliation[e]{Theory Division, CERN, CH-1211 Geneva 23, Switzerland}
\affiliation[f]{University of Alberta, Edmonton AB T6G 2J1, Canada}
\affiliation[g]{Institut f{\"u}r Theoretische Teilchenphysik, KIT, 76128 Karlsruhe, Germany}
\affiliation[h]{Department of Physics, University of Arizona, Tucson AZ 85721, USA}
\affiliation[i]{IFIC (UVEG/CSIC), Ap. Correos 22085, E-46071 Valencia, Spain}
\affiliation[j]{Department of Physics \& Astronomy, University of California - Riverside, Riverside, USA}

\note{Preprint numbers: CERN-PH-TH/2013-226}

\abstract{We present an up-to-date overview of the problem of top quark mass determination. We assess the need for precision in the top mass extraction in the LHC era together with the main theoretical and experimental issues arising in precision top mass determination. We collect and document existing results on top mass determination at hadron colliders and map the prospects for future precision top mass determination at $e^+e^-$ colliders. We present a collection of estimates for the ultimate precision of various methods for top quark mass extraction at the LHC.}

\begin{document} 
\maketitle
\flushbottom

\section{Introduction}

The precision with which we determine the top quark mass impacts our understanding of several phenomena.
Examples are EW precision fits \cite{Baak:2012kk}, determination of the vacuum stability in the Standard Model \cite{Degrassi:2012ry,Bezrukov:2012sa} as well as models with broad cosmological implications \cite{Bezrukov:2007ep,DeSimone:2008ei}. 

A number of measurements of $m_t$ from hadron colliders exist~\cite{TEV:2012mt,LHC:2012mt}, utilizing all measured decay modes of the top quark. The experimental extraction has accuracy of $\delta m_t \lesssim 1\, {\rm GeV}$. The main task for this writeup is to map the steps that can clarify the relation between the extracted value of the mass and a theoretically well-defined top mass (like the pole mass). 

The top quark mass $m_t$ is not a physical observable and, therefore, it cannot be measured directly. Virtually all existing strategies for determining $m_t$ (see section~\ref{sec:methods}) are based on its extraction from observables that are directly
sensitive to it, i.e. $m_t$ is defined as the solution to the following implicit relation: 
\begin{equation}
\label{exptheory}
\sigma^{\rm exp}(\{Q\})  = \sigma^{\rm th}(m_t, \{Q\})   \, ,
\end{equation}
where $\{Q\}$ is a set of kinematical variables, $\sigma^{\rm exp}$
stands for the measured and $\sigma^{\rm th}$ for the predicted value
of some chosen observable $\sigma$.  In a typical application, $m_t$ is adjusted in $\sigma^{\rm th}$ to obtain the best fit to the shape of $\sigma^{\rm exp}$, as a function of the variables $\{ Q\}$. This implicitly assumes that $\sigma^{\rm exp}$ has been corrected for detector (and possibly acceptance) effects, or that the converse has been applied to $\sigma^{\rm th}$, so that the observables on either side of Eq.~\eqref{exptheory} are defined at the same level, with the same cuts.
Uncertainties in the theoretical prediction due to missing higher-order effects, finite-width effects,  and non-perturbative corrections are generally present. We discuss them in more detail in section~\ref{sec:issues}.

The top mass $m_t$ is scheme dependent and a large number of such schemes exist. Examples are the pole, the $\overline{\rm MS}$, and the 1S schemes \cite{Hoang:1999zc}; see Ref.~\cite{Hoang:2008xm,Ahrens:2011px} for a discussion in the context of hadron colliders. Different mass schemes are perturbatively related to each other. For example,  the top mass $m_t(R,\mu)$ in scheme ``$R$" is related to the pole mass $m_t^{\rm pole}$ through a perturbative series
\begin{equation}
m_t^{\text{pole}} = m_t(R,\mu) + \delta m_t(R,\mu), \qquad \delta m_t(R,\mu) = R\sum_{n=1}^{\infty} \sum_{k=0}^{n} a_{nk} \left[ \alpha_s(\mu)\right]^n \ln^k \left(\frac{\mu^2}{R^2} \right),
\end{equation}
where $R$ is a scale associated with the scheme; for the $\overline{\rm MS}$ scheme $R\sim m_t$. The relation between the pole and $\overline{\rm MS}$ masses is known to three loops in QCD \cite{Chetyrkin:1999ys,Melnikov:2000qh}; a possible large EW correction has recently been reported in Ref.~\cite{Jegerlehner:2012kn}. Large logarithms can arise in converting between schemes if the scale $R\ll m_t$, as seen in top resonance schemes \cite{Fleming:2007xt} where $R\sim \Gamma_{\rm top}$, and can be resummed via an infrared renormalization group equation~\cite{Hoang:2008yj}. 

A reliable interpretation of top mass measurements requires understanding the connection between the theory prediction, in a given top mass scheme, and the experimental observable as shown  schematically in Eq.(\ref{exptheory}). This  connection between experimental observables and the appropriate top mass schemes is well understood in $e^+e^-$ colliders. Precision top quark mass determinations at $e^+e^-$ colliders have been studied for top pair production near threshold~\cite{Fadin:1987wz,Strassler:1990nw,Fadin:1988fn,Fadin:1991zw} and in the boosted regime \cite{Fleming:2007qr,Fleming:2007xt}. The expected uncertainty in the top mass from the threshold scan method is $\delta m_t \lesssim 100\, {\rm MeV}$~\cite{Penin:2012pd,Hoang:2000yr} and a few hundred MeV~\cite{Chekanov:2003cp} for boosted top quarks. See sections~\ref{sec:e+e-basics},\ref{sec:e+e-theory},\ref{sec:e+e-exp} for details. For hadron colliders, which are the main focus of the current and near future research, and by extension of this document, the situation is more complicated and a rigorous framework is still lacking. Below we review the current status and issues in precision top mass extractions at hadron colliders.

\section{Issues in precision top mass determination at hadron colliders}\label{sec:issues}

A unique property of the top quark is that it decays very quickly, before it can form strongly interacting bound states. For this reason the top quark can be studied largely free of non-perturbative effects \cite{Fadin:1987wz,Strassler:1990nw,Fadin:1988fn}. Still, a number of uncertainties of perturbative and non-perturbative origin affect the extraction of $m_t$:
\begin{enumerate}
\item {\it MC modeling.} Most methods for 
  extraction of $m_t$ rely on modeling the measured final
  state with typically LO+LL MC generators. 
The extracted mass then reflects the mass parameter in the corresponding MC
generator. Identifying the nature of this mass parameter and relating
it to common mass schemes, like the pole mass, is a non-trivial and
open problem, and may be associated with ambiguities of order 1
GeV~\cite[Appendix C]{Buckley:2011ms}; see also Ref.~\cite{Hoang:2008xm}. 
The effect of the top and
bottom masses on 
parton-shower radiation patterns is generally included already in the
LO+LL Monte Carlos~\cite{Corcella:1998rs,Norrbin:2000uu,Hamilton:2006ms,Mangano:2006rw,Schumann:2007mg,Alwall:2008qv}
and acts to screen the collinear singularities. NLO matching and
non-perturbative effects are discussed separately below. 
\item {\it Reconstruction of the top pair.} 
  Typically, the existing methods for extraction of the top quark mass
  implicitly or explicitly rely on the reconstruction of the top pair
  from final state leptons and jets. This introduces uncertainties of
  both perturbative origin (through higher-order corrections) and
  non-perturbative origin (related to hadronization and non-factorizable
  corrections). Methods that do not rely on such reconstruction are
  therefore complementary and highly desirable; two examples are given 
  in \ref{sec:methods}.\ref{JPsi} and \ref{sec:methods}.\ref{dilepton}.
\item {\it Unstable top and finite top width effects.} These effects have
been studied extensively in the context of top pair production at $e^+e^-$
colliders \cite{Hoang:2010gu,Beneke:2010mp,Penin:2011gg}. In the context of
higher order corrections at hadron colliders, finite top (and $W$) width
effects have been computed in \cite{Denner:2012yc,Bevilacqua:2010qb} where
comparisons versus the narrow width approximation can be found. The
conclusion is that these corrections are small, sub-1\%, in inclusive
observables (like the total inclusive cross-section used in
\ref{sec:methods}.\ref{sigmatot}) but can be sizable in tails of kinematical
distributions. In particular, they significantly affect the tail of the $B\,
\ell$ invariant mass distribution used in the method
\ref{sec:methods}.\ref{JPsi} (but not the central region of the distribution
which is most relevant for the $m_t$ determination described in
\ref{sec:methods}.\ref{JPsi}).
\item {\it Bound-state effects in top pair production at hadron colliders.}
The effect of bound state formation on top pair production at hadron
colliders has been studied in
Refs.~\cite{Fadin:1990wx,Hagiwara:2008df,Kiyo:2008bv,Sumino:2010bv}. It can
dramatically affect the shape of differential distributions within a few GeV
of absolute threshold. Therefore, any mass measurement that is sensitive to
this kinematical region has to properly take these effects into
consideration. In the context of the total cross-section, see
Refs.~\cite{Bonciani:1998vc,Beneke:2011mq}, the effect on the cross-section
is sub-1\% and is taken into account in current higher order calculations of
the total inclusive cross-section (and thus in mass extractions based upon
it).
\item {\it Renormalon ambiguity in top mass definition.}  It is well known \cite{Bigi:1994em,Beneke:1994sw,Smith:1996xz,Hoang:2008yj} that the pole mass of the top quark suffers from the so-called renormalon ambiguity. This implies an additional irreducible uncertainty of several hundred MeV's on the top pole mass. The short distance masses do not suffer from the renormalon ambiguity and the precision in their determination is restricted only by experimental and theoretical uncertainties. At hadron colliders, where currently $\delta m_t \lesssim 1\, {\rm GeV}$, the renormalon ambiguity is numerically subdominant; see also Ref.~\cite{Ahrens:2011px}.
\item {\it Alternative top mass definitions.} It is well understood from $e^+e^-$ collider studies that by using alternative top mass definitions one could improve the precision of the extracted top quark mass. Similar studies for hadron colliders have been done in Refs.~\cite{Hoang:2008xm,Langenfeld:2009wd,Ahrens:2011px}. It has been argued in Ref.~\cite{Hoang:2008xm} that for top mass extractions in the peak region, the appropriate short distance mass schemes correspond to the top resonance schemes where $R \sim R_{sc} \sim 1$ GeV $\sim \Gamma_{\rm top}$ , where $R_{sc}$ is the shower cutoff implemented in the MC. An interpretation of this statement in the context of a factorization framework for hadron colliders is still lacking. Ref.~\cite{Langenfeld:2009wd} advocates extracting directly the top $\overline{\rm MS}$ mass from the top pair production cross-section. The improvement at the current level of precision $\delta m_t \lesssim 1\, {\rm GeV}$, however, is small \cite{Ahrens:2011px} (see also the discussion about renormalon ambiguity, above). 
The extracted top $\overline{\rm MS}$ mass might be affected by the findings reported in Ref.~\cite{Jegerlehner:2012kn}.
\item {\it Higher-order corrections.} Missing higher-order corrections
  can be an 
important source of uncertainty in the determination of the top mass. 
These are typically added through NLO calculations
\cite{Melnikov:2009dn,Biswas:2010sa,Denner:2012yc,Bevilacqua:2010qb}
and for the case of the total cross-section through approximate NNLO calculations
\cite{Langenfeld:2009wd,Ahrens:2011px,Beneke:2012wb,Beneke:2011mq} (for calculations in full NNLO, see the discussion in \ref{sec:methods}.\ref{sigmatot} below). A particularly sensitive issue is the matching of NLO top-quark
calculations to parton
showers, see~\cite{Frixione:2003ei,Frixione:2007nw,Nason:2012pr}. 
\item {\it Non-perturbative corrections.} 
Non-perturbative corrections mostly affect the MC modeling of the
final state. These include hadronization, 
in particular of the final-state partons that
inherit the top quark color charges (which causes an unavoidable 
non-perturbative exchange of energy with the rest of the event), 
hadron and $\tau$ decays (including the $B$ hadron decays), 
underlying event, and possible additional non-perturbative phenomena
such as color reconnections or other collective phenomena. Depending
on how the corrections to the
cross-sections in eq.~\eqref{exptheory} are performed, these
uncertainties enter either on the experimental or theoretical side
of the equation. The underlying-event, hadronization, and particle-decay 
corrections are typically dealt with at the jet-calibration stage, and
the resulting systematic uncertainties become part of the
jet-energy-scale (JES) systematics. 
A study of color-reconnection effects in the special
case of $e^+e^-$ collisions found very small effects $<$ 100
MeV~\cite{Khoze:1994fu}, but toy models show that 
the effect in hadron collisions may be
as large as 0.5~GeV~\cite{Skands:2007zg}. More physical models and
better constraints are required to reduce this uncertainty further, 
for instance by allowing one to bound it, rather than 
merely switching it on and off.
Non-perturbative corrections  
can also be introduced through final-state interactions in the
presence of strong jet vetoes \cite{Mitov:2012gt}. Inclusive
measurements like the methods described in sections \ref{sec:methods}.\ref{JPsi} and \ref{sec:methods}.\ref{dilepton} are likely to suffer least from such non-perturbative effects.  

\item {\it Contributions from physics beyond the Standard Model.} 
It is possible that some yet-undiscovered physics beyond the Standard Model (BSM) might influence the various measurements used to extract the top quark mass. Given that in the context of top mass extraction experimental measurements have so far always been compared with predictions based on the SM, the possibility arises that there might be a bias in the determination of the top quark mass due to new physics. While it is unlikely that such new physics can cause large corrections,
\footnote{For example, the CMS end-point top mass determination (see sec.~\ref{sec:CMS-endpoint}) is based on kinematical considerations, i.e. it has reduced sensitivity to the top quark production mechanism.}
${\cal O}(1~{\rm GeV})$ modifications to $m_t$ cannot be excluded at present. A first dedicated study of BSM contributions to $m_t$ determination is ongoing \cite{CMPRW}. Application to top mass measurements of the work reported in Ref.~\cite{Agashe:2013eba} may also be useful for disentangling BSM contributions (although this will likely require the inclusion of NLO QCD corrections). 

\end{enumerate}

\section{Top mass determination at hadron colliders}\label{sec:methods}

A major collection of experimental methods is available in \cite{Galtieri:2011yd}.  Here we highlight a few that have already proven useful or appear to be promising:

\begin{enumerate}
\item\label{MEM} {\it Matrix element methods}. The most precise measurements of $m_t$ from the Tevatron use the matrix element method~\cite{Abazov:2004me,Estrada:2001me}, in which the measured objects are compared with expectations from the LO $t\bar{t}$ production and decay diagrams convoluted with the detector response.   The method derives much of its power from the fact that the likelihood for each event to be consistent with both $t\bar{t}$ and background production is calculated; greater weight is assigned to events that are more likely to be from $t\bar{t}$ when measuring $m_t$.  In addition, the hadronically-decaying $W$ boson  in $\ell +$ jets events provides an {\it in situ} constraint on the jet response, substantially reducing the systematic uncertainty.  An NLO theory approach is currently being developed \cite{Campbell:2012cz}. 
\item\label{ideogram} {\it Ideogram and template methods}. The current generation of CMS analyses, which are among the most precise $m_t$ measurements, use Ideogram techniques. The ideogram corresponding to the most probable solution for the mass is
determined on an event-by-event basis. These are then summed over the full
dataset to determine an ``integrated ideogram". The top
mass is then determined by fitting this to a Monte Carlo spectrum for the
same number of events. The MC spectrum is determined as a function of $m_t$
(CMS - all jets) or $m_t$ and JES (CMS - lepton and jets). The dilepton channel is
handled in a similar way using the analytical matrix weighting technique (AMWT) to treat the 2-neutrino
ambiguities. Regarding Monte Carlo generators, CMS uses Madrgraph (LO ME generator) with Pythia for the parton showering.

The ATLAS collaboration uses similar ``template" methods. The main differences with respect to the CMS analyses are that the ATLAS Collaboration currently uses 3-parameters ($m_t$, lightJES, bJES) for their lepton + jets analysis as well as MC@NLO + Herwig for event generation. 
\item\label{sigmatot} {\it Extraction from the total cross-section
  $\sigma_{\rm tot}$}. The total inclusive $t\bar{t}$ cross-section at a given collider depends on $m_t$, so the measured cross-section can be used to constrain $m_t$. 
Extractions of the top mass from $\sigma_{\rm tot}$ 
have been performed in \cite{Langenfeld:2009wd,Ahrens:2011px,Beneke:2012wb,Beneke:2011mq} using NLO+NNLL or approximate NNLO cross-section calculations.  Very recently a first analysis performed in full NNLO+NNLL appeared \cite{Chatrchyan:2013haa}. The sensitivity of $\sigma_{\rm tot}$ to the top mass is relatively low (few \%), so this method is not competitive in precision with other existing methods. On the other hand the method uses an observable based on a well-defined top mass, has small uncertainties due to perturbative and non-perturbative effects, and is not very sensitive to top width effects. 
\item\label{JPsi} {\it The $J/\psi$ method} \cite{Kharchilava:1999yj}. 
In about one in $10^5$ top quark decays, the fragmentation products of the $b$ quark  will include 
a $J/\psi$ decaying to $\mu^+\mu^-$.  If the $W$ boson from the same top quark also decays leptonically, the three-lepton invariant mass is sensitive to $m_t$.  
The other top quark is only used to discriminate $t\bar{t}$ production from background.  The strength of this method is that the main systematic uncertainties arise from different sources than in other methods (primarily $b$ fragmentation), and may be smaller. Moreover, no $t\bar t$ reconstruction takes place i.e. the method is inclusive at any order in perturbation theory. These potential advantages must be weighed against the statistical limitations arising from requiring a $J/\psi$ candidate.
MC studies of this method are reported in \cite{Corcella:2000wq,Chierici:topmass}, and the 
uncertainty from $b$ fragmentation was studied at NLO in
\cite{Corcella:2001hz,Cacciari:2002re,Corcella:2005dk}. A NLO
study, with factorized production and decay, was performed in
Ref.~\cite{Biswas:2010sa}. The complete NLO result including
production/decay interferences, off-shell effects and backgrounds, was
computed in Ref.~\cite{Denner:2012yc} (the $B$ mesons in this work are
treated as $b$-jets). Additional error estimates, performed within this study, can be found in sec.~\ref{sec:JPsi-method} below.
\item\label{dilepton} {\it Dilepton-specific methods}. In the same spirit as the $J/\psi$ method, it may be advantageous to measure $m_t$ using kinematic properties (e.g. the invariant mass and $p_T$) of the lepton pair  in dilepton $t\bar{t}$ candidates (selected as pair of leptons and possibly two $b$'s, without requiring $t\bar t$ reconstruction) \cite{top-mass-1}.  These observables should have a smaller sensitivity to the modeling of hadronic observables (showering and jets). Such measurements can be compared versus complete NLO calculations \cite{Denner:2012yc,Bevilacqua:2010qb}, as well as versus standard MC generators. This approach may not be as sensitive to the value of $m_t$ as other methods, but offers very different systematics, and therefore may help to reduce the overall uncertainty on the world-average $m_t$. First measurements of top pair differential distributions in dilepton final states have already appeared \cite{Chatrchyan:2012saa}. See also the related discussion in sec.~\ref{sec:kin-distrib} below.
\end{enumerate}

In the near to medium term (i.e. prior to the construction of a lepton collider capable of performing a $t\bar{t}$ threshold scan), improvement in the precision with which we know $m_t$ will depend on:
\begin{itemize}
\item Extraction of the top mass with new methods that have alternative systematics (like \ref{JPsi} and \ref{dilepton} in section~\ref{sec:methods}). Such extractions will either validate the current precision in the available top mass measurements or highlight the need for additional scrutiny. Further phenomenological and experiment studies of these new methods are needed.
\item Decreasing the perturbative uncertainty in currently used Matrix Element methods by applying future extension of the work in Ref.~\cite{Campbell:2012cz}. It remains an open question if top width effects and non-perturbative effects can also be reduced this way.
\item Improved understanding of the relation between MC mass and
  standard quark masses, such as the pole mass. Work along these lines
  has been reported in \cite{Hoang:2008xm}; see also
  Ref.~\cite[Appendix C]{Buckley:2011ms}. 
\end{itemize}

In the following we review, and present estimates, for the capabilities of various methods for top mass determination. The methods can be split into ``conventional" (sec.~\ref{sec:conventional}), ``other available" (sec.~\ref{sec:CMS-endpoint}, \ref{sec:ATLAS-3d-fit}) or ``under development" (sec.~\ref{sec:JPsi-method}, \ref{sec:kin-distrib}).

\subsection{``Conventional" top mass determination techniques}\label{sec:conventional}

As a model for the conventional collider mass measurements, we consider the CMS lepton-plus-jets \cite{Chatrchyan:2012cz}, dilepton \cite{Chatrchyan:2012ea} and all-hadronic analyses \cite{Chatrchyan:2013xza}. These are currently the most precise measurements in each channel. The analyses use similar methods and result in measurements with comparable systematic uncertainties. To estimate the potential precision for the various 14 TeV scenarios we have taken the CMS lepton-plus-jet result $m_t = 173.49 \pm 0.27 ({\rm stat.}) \pm 1.03 ({\rm syst.})$ GeV as representative and have performed extrapolations based on this. The results are presented in Table~\ref{l+jet}.
\begin{table}[h]
\begin{center}
\begin{tabular}{|c|c|c|c|c|c|c|c|c|}
\hline
& {\rm Ref.}\cite{Chatrchyan:2012cz} & \multicolumn{5}{c|} {\rm Projections}\\
\hline
{\rm CM Energy} & {\rm 7 TeV } & \multicolumn{5}{c|}{\rm 14 TeV  } \\
\hline
{\rm Cross Section} & {\rm 167 pb} & \multicolumn{5}{c|}{\rm 951 pb}  \\
\hline
{\rm Luminosity} & {$5 fb^{-1}$} & \multicolumn{2}{c|}{$100 fb^{-1}$} & \multicolumn{2}{c|}{$300 fb^{-1}$} & {$3000 fb^{-1}$} \\ 
\hline
{\rm Pileup} & 9.3 & 19 & 30 & 19 & 30  & 95  \\
\hline\hline
{\rm Syst. (GeV)} & 0.95 & 0.7 & 0.7 & 0.6 & 0.6 & 0.6 \\
\hline
{\rm Stat. (GeV)} & 0.43 & 0.04 & 0.04 & 0.03 & 0.03 & 0.01 \\
\hline\hline
{\bf Total } & {\bf 1.04} & {\bf 0.7} & {\bf 0.7} & {\bf 0.6} & {\bf 0.6} & {\bf 0.6} \\
\hline
{\rm Total} (\%) & 0.6 & 0.4 & 0.4 & 0.3 & 0.3 & 0.3 \\
\hline
\end{tabular}
\caption{Extrapolations based on the published CMS lepton-plus-jets analysis}
\label{l+jet}
\end{center}
\end{table}

These are based on the 7 and 14 TeV cross-sections calculated using the full NNLO framework \cite{Czakon:2013goa} with an allowance for a decreased trigger efficiency due to higher event rates and trigger thresholds.  For the systematic errors, we assume that some of the soft QCD and fragmentation uncertainties will be constrained using the data from future LHC runs.  We keep the initial and final state radiation and pdf uncertainties unchanged. Without a full simulation of the machine conditions, we are unable to model the effects of the increased merging of the top-decay products in moving to the higher energy. To allow for this and the uncertainties in the extrapolations we add in an additional 300 MeV uncertainty to the mass measurement. 

\begin{table}[h]
\begin{center}
\begin{tabular}{|c|c|}
\hline
{\rm Scenario} & {\rm Dominant Uncertainties} \\
\hline
{\rm Ref.}\cite{Chatrchyan:2012cz} & {\rm Jet Energy Scale, Hadronization, Soft QCD, ISR/FSR}\\
\hline
{\rm 100 $fb^{-1}$/19 PU} & {\rm Jet Energy Scale, Hadronization, Soft QCD, ISR/FSR} \\
\hline
{\rm 100 $fb^{-1}$/30 PU} & {\rm Jet Energy Scale, Hadronization, Soft QCD, ISR/FSR, Pileup} \\
\hline
{\rm 300 $fb^{-1}$/19 PU} & {\rm Jet Energy Scale, Hadronization, Soft QCD, ISR/FSR} \\
\hline
{\rm 300 $fb^{-1}$/30 PU} & {\rm Jet Energy Scale, Hadronization, Soft QCD, ISR/FSR, PIleup} \\
\hline
{\rm 3000 $fb^{-1}$/95 PU} & {\rm Jet Energy Scale, Hadronization, Soft QCD, ISR/FSR, PIleup } \\
\hline
\end{tabular}
\caption{Dominant systemic uncertainties for each scenario}
\label{l+jet big error}
\end{center}
\end{table}

In Table~\ref{l+jet big error} we summarize the dominant uncertainties for each scenario. While these are very similar, it should be noted that pileup and the associated uncertainties from the missing transverse energy and contamination of the underlying event are expected to become increasingly important as the collision energy and pileup are increased. We also note that the ISR/FSR uncertainly, that is one of the sub-leading uncertainties for  \cite{Chatrchyan:2012cz} becomes one of the leading uncertainties for each of the 300 $fb^{-1}$ and 3000 $fb^{-1}$ scenarios.

Based on the comparison of the results from  \cite{Chatrchyan:2012cz} and the CMS combined result from the three channels shown at the TOP2012 Workshop \cite{TOP2012}, see also \cite{combination}, we estimate that combinations of different channels for each of the 14 TeV scenarios may lead to a small improvement in the projected precisions. We also note that the triggering on the all-hadronic events may prove difficult when running at very high luminosity and under high pileup conditions. This may prevent the effective use of this channel under these conditions.

\subsection{CMS end-point method \cite{Chatrchyan:2013boa}}\label{sec:CMS-endpoint}

This method is kinematical in nature and utilizes the correlation between the end-points of the $M_{b\,\ell}$ and the $M_{T2perp}^{221}$ distributions and $m_t$. It gives a mass measurement $m_t = 173.90 \pm 0.90 ({\rm stat.}) ^{+1.70}_{-2.1} ({\rm syst.})$ GeV. This was extrapolated using similar assumptions to that used for the CMS lepton-plus-jet method. A summary of the results is given in table~\ref{endpoint}. As this technique is insensitive to pileup effects we only quote one extrapolation for each of the luminosity scenarios.

\begin{table}[h]
\begin{center}
\begin{tabular}{|c|c|c|c|c|c|c|c|c|}
\hline
& {\rm Ref.}\cite{Chatrchyan:2013boa} & \multicolumn{3}{c|} {\rm Projections}\\
\hline
{\rm CM Energy} & {\rm 7 TeV } & \multicolumn{3}{c|}{\rm 14 TeV  } \\
\hline
{\rm Cross Section} & {\rm 167 pb} & \multicolumn{3}{c|}{\rm 951 pb}  \\
\hline
{\rm Luminosity} & {$5 fb^{-1}$} & {$100 fb^{-1}$} & {$300 fb^{-1}$} & {$3000 fb^{-1}$} \\ 
\hline\hline
{\rm Syst. (GeV)} & 1.8 & 1.0 & 0.7 & 0.5 \\
\hline
{\rm Stat. (GeV)} & 0.90 & 0.10 & 0.05 & 0.02 \\
\hline\hline
{\bf Total } & {\bf 2.0} & {\bf 1.0} & {\bf 0.7} & {\bf 0.5} \\
\hline
{\rm Total} (\%) & 1.2 & 0.6 & 0.4 & 0.3 \\
\hline
\end{tabular}
\caption{Extrapolations based on the published CMS Endpoint analysis}
\label{endpoint}
\end{center}
\end{table}

In Table ~\ref{Endpoint big error} we summarize the dominant uncertainties for each scenario. As with the conventional analysis, these are fairly similar as a function of increasing luminosity. We also note that, unlike the conventional method, the ISR/FSR and pileup terms do not seem to play a role in the precision of the measurements, even at high luminosity.

\begin{table}[h]
\begin{center}
\begin{tabular}{|c|c|}
\hline
{\rm Scenario} & {\rm Dominant Uncertainties} \\
\hline
{\rm Ref.}\cite{Chatrchyan:2013boa} & {\rm Jet Energy Scale, Hadronization, Soft QCD}\\
\hline
{\rm 100 $fb^{-1}$} & {\rm Jet Energy Scale, Hadronization, Soft QCD} \\
\hline
{\rm 300 $fb^{-1}$} & {\rm Jet Energy Scale, Hadronization, Soft QCD} \\
\hline
{\rm 3000 $fb^{-1}$} & {\rm Jet Energy Scale, Hadronization} \\
\hline
\end{tabular}
\caption{Dominant systemic uncertainties for each scenario}
\label{Endpoint big error}
\end{center}
\end{table}

Although the terms listed in Tables ~\ref{l+jet big error} and ~\ref{Endpoint big error} have a large overlap, we  note that they are not 100\% correlated so that combining the results from the two methods may be beneficial to the overall precision. This follows from the fact that, unlike the conventional analyses, the Endpoint method does not rely on Monte Carlo modeling to do an internal calibration. It is largely analytical with a data-driven model for the background.

We also note that the kinematical nature of this method makes it suitable to attempt top mass determination which is less likely to be affected by possible new physics contributions. Nonetheless, this important aspect of $m_t$ determination needs further study. Finally, one would like to study in more detail the effect of higher order corrections, for example, by comparing with the findings of Refs.~\cite{Bevilacqua:2010qb,Denner:2012yc}.

\subsection{ATLAS 3-dimensional template fit method \cite{ATLAS-3-dim-fit}}\label{sec:ATLAS-3d-fit}

The ATLAS collaboration has recently published a new determination of the 
top quark mass in the lepton+jets final state~\cite{ATLAS-3-dim-fit}.
This analysis uses a 3-dimensional template technique which determines 
the top quark mass together with two important experimental systematic 
uncertainties.

The result is $m_t =$ 172.31 $\pm$ 0.23 (stat) $\pm$ 0.27 (JSF) $\pm$ 
0.67 (bJSF) $\pm$ 1.35 (syst) GeV. The uncertainties labeled JSR and 
bJSF correspond to the {\em statistical} uncertainty of the global jet energy 
scale factor (JSF) and the relative b-jet to light-jet energy scale factor 
(bJSF). The in-situ determination of these uncertainties in the 3D fit 
has allowed the two dominant systematic uncertainties to be transformed into 
statistical uncertainties to a large extent. 
The residual Jet Energy Scale uncertainty is combined together with
a large number of other sources of uncertainty into ``syst". 
The modeling of top quark production and decay has a 
non-negligible contribution.

\subsection{Top mass determination from J/$\Psi$ final states \cite{Kharchilava:1999yj}}\label{sec:JPsi-method}

Our estimate of the theory error is based on the NLO QCD calculation of Ref.~\cite{Biswas:2010sa} performed for LHC 14 TeV.  The estimation of the statistical uncertainty is based on preliminary studies by the CMS collaboration. Calculations for LHC 33 TeV in leading order QCD are also available.
\footnote{We thank the authors of Ref.~\cite{Biswas:2010sa} for providing us with these additional estimates.}
From these results we conclude that $\langle M_{B\ell}\rangle( m_t )$ is not sensitive to the collider energy, if the same cuts are used. More restrictive cuts for LHC 33 TeV lead to slight modification of the $\langle M_{B\ell}\rangle( m_t )$ dependence, but the theoretical error of the extracted $m_t$ remains largely unchanged. 

The main sources of theoretical error in the J/$\Psi$ method are scale variation and $B$-fragmentation. Modeling of $\langle M_{B\ell}\rangle$ in NNLO QCD could become possible during the LHC 13 TeV run, which would reduce the scale variation by a factor of 2.5. We estimate this possible improvement by comparing in Table~\ref{tot-xsec-13-33} the scale and pdf uncertainty of the total inclusive cross-section for LHC 13 and 33 TeV at NLO and NNLO \cite{Czakon:2013goa}.
\begin{table}[h]
\begin{center}
\begin{tabular}{|c|c|c|c|c|c|c|c|c|}
\hline
& \multicolumn{4}{|c|}{LHC 13 TeV} & \multicolumn{4}{|c|}{LHC 33 TeV}\\
\hline
& \multicolumn{2}{|c|}{$\delta_{\rm scale} [\%]$} & \multicolumn{2}{|c|}{$\delta_{\rm pdf}[\%]$} & \multicolumn{2}{|c|}{$\delta_{\rm scale}[\%]$} & \multicolumn{2}{|c|}{$\delta_{\rm pdf}[\%]$} \\
\hline
& MSTW & NNPDF & MSTW & NNPDF & MSTW & NNPDF & MSTW & NNPDF \\
\hline\hline
{\rm NLO} & $ ^{+12.1}_{-12.1} $ &$ ^{+11.8}_{-11.9} $ & $ ^{+1.9}_{-2.3}$&$ ^{+1.8}_{-1.8} $ & $ ^{+11.5}_{-10.3}$&$ ^{+11.2}_{-10.0} $ & $ ^{+1.2}_{-1.5}$ &$ ^{+1.0}_{-1.0} $ \\
\hline
{\rm NNLO} & $ ^{+3.4}_{-5.6} $&$ ^{+3.5}_{-5.7} $ & $ ^{+1.8}_{-2.0}$&$ ^{+1.8}_{-1.8} $ & $ ^{+3.1}_{-4.7}$&$ ^{+3.1}_{-4.7} $ & $ ^{+1.0}_{-1.4}$&$ ^{+1.0}_{-1.0} $ \\
\hline
\end{tabular}
\caption{Scale and pdf uncertainty for the total inclusive $t\bar t$ cross-section at 13 and 33 TeV. }
\label{tot-xsec-13-33}
\end{center}
\end{table}
We use $m=173.3$ GeV with MSTW2008 \cite{Martin:2009iq} (with 68cl) and NNPDF2.3~\cite{Ball:2012cx} (with $\alpha_s(M_Z)=0.118$ and $n_f=5$) NLO and NNLO pdf sets.

The long-term limiting factor would be the uncertainty in $B$-fragmentation. As a benchmark, we take the DELPHI measurement \cite{DELPHI} of the first moment of the fragmentation function $\langle x \rangle = 0.7153 \pm 0.0052$, which has an uncertainty of about $0.7\%$ (completely dominated by systematics). Such error in $\langle M_{B\ell}\rangle$ implies $\delta m_t \approx 0.9$ GeV. A future dedicated ILC run at the $Z$-pole should be able to improve this measurement significantly. Such a measurement is likely to occur only after the end of the currently foreseen LHC operations and before the dedicated top threshold scan during the later phases of the ILC where, for the first time, measurement of $m_t$ with very high precision ${\cal O}(100~ {\rm MeV})$ will be performed (see sections~\ref{sec:e+e-basics}, \ref{sec:e+e-theory}, \ref{sec:e+e-exp}).

The estimates for the total error are given in Table~\ref{table:JPsi}. The theoretical error is estimated as follows: for LHC 8 and 14 TeV and luminosity up to $300 fb^{-1}$ we take the error as estimated in Ref.~\cite{Biswas:2010sa}. For $3000 fb^{-1}$ at 14 TeV we assume that NNLO calculation will be available, which will decrease the scale uncertainty by a factor of 2.5. At this point the dominant uncertainty is the one from $B$-fragmentation. For LHC at 100TeV we assume that the $B$-fragmentation uncertainty is reduced by a factor of 2 with the help of a dedicated future lepton collider measurement.
\begin{table}[h]
\begin{center}
\begin{tabular}{|c|c|c|c|c|c|c|c|c|}
\hline
& {\rm Ref. analysis} & \multicolumn{5}{|c|}{\rm Projections} \\
\hline
{\rm CM Energy} & {\rm 8 TeV } & \multicolumn{3}{c|}{\rm 14 TeV  } & {\rm 33 TeV} & {\rm 100 TeV}\\
\hline
{\rm Cross Section} & {\rm 240 pb} & \multicolumn{3}{c|}{\rm 951 pb}  & {\rm  5522 pb} & {\rm 25562 pb}\\
\hline
{\rm Luminosity} & {$20 fb^{-1}$} & {$100 fb^{-1}$} & {$300 fb^{-1}$} & {$3000 fb^{-1}$} & {$3000 fb^{-1}$} & {$3000 fb^{-1}$} \\ 
\hline\hline
{\rm Theory (GeV)} & - & 1.5 & 1.5 & 1.0 & 1.0 & 0.6 \\
\hline
{\rm Stat. (GeV)} & 7.00 & 1.8 & 1.0 & 0.3 & 0.1 & 0.1 \\
\hline
{\bf Total} & {\bf -} & {\bf 2.3} & {\bf 1.8} & {\bf 1.1} & {\bf 1.0} & {\bf 0.6}\\
\hline
{\bf Total (\%)} & {\bf -} & {\bf 1.3} & {\bf 1.0} & {\bf 0.6} & {\bf 0.6} & {\bf 0.4}\\
\hline\hline
\end{tabular}
\caption{Extrapolations based on the $J/\Psi$ method.}
\label{table:JPsi}
\end{center}
\end{table}

\subsection{Top mass determination from kinematic distributions}\label{sec:kin-distrib}

The top quark mass can be extracted from $\sigma_{\rm tot}$. The advantage
of this method is that a mass is obtained in a rigorously defined mass scheme.
The D0 experiment has attempted this approach~\cite{Abazov:2011pta}. 
Preliminary results have been presented by both the ATLAS and CMS Collaborations. The uncertainty
on the extracted top quark mass amounts to approximately 3\%. 
Although the recently derived NNLO result \cite{Czakon:2013goa} has not yet 
been fully utilized in this regard (however see Ref.~\cite{Chatrchyan:2013haa}), significant future improvements 
within this approach are unlikely given that the uncertainty in 
$\sigma_{\rm tot}$ at present arises from a number of competing sources 
\cite{Czakon:2013tha}. Ultimately the potential of this method is 
expected to be limited by the relatively small sensitivity of the cross 
section with respect to the top quark mass.

Kinematic differential distributions offer improved sensitivity to $m_t$. 
Ref.~\cite{Alioli:2013mxa} suggested $m_t$ extraction from the 
invariant mass distribution of $t\bar t$ pairs produced in events in 
association with a hard jet. The sensitivity is improved well beyond 
what can be achieved with the total cross-section. The authors claim 
that uncertainties related to uncalculated higher order
corrections or uncertainties in the parton distribution functions
are expected to affect the mass measurement by less than 1 GeV. 
The impact of top decays and experimental uncertainties - evaluated in a 
generic detector simulation - is also expected to be sub-GeV.

The extraction of $m_t$ from leptonic kinematic distributions in dilepton events \cite{top-mass-1} is less affected by MC modeling and non-perturbative corrections, thus reducing an important source of uncertainty in the current top mass extractions. The only currently available study of $m_t$ extraction from dilepton events has been performed for LHC 14 TeV in Ref.~\cite{Biswas:2010sa} where the authors find the possibility for extracting $m_t$ with precision of about 1.5 GeV. Such a precision is similar to the one from the J$/\Psi$ method. Further exploration of the systematics in this method is needed and studies are currently underway \cite{top-mass-1}.

\section{Top mass determination at lepton colliders}\label{sec:e+e-basics}

Current theoretical understanding of top quark threshold production at lepton colliders suggests (see sec.~\ref{sec:e+e-theory} below) that it is feasible to determine the top quark mass with a precision of about 100 MeV, the top quark width with a precision of about 40 MeV and the top quark Yukawa coupling with a precision of about 50\%. Such a precision is substantially higher than the ultimate precision expected at hadron colliders.

Several proposals for lepton colliders -- mainly linear $e^+e^-$ colliders -- have been put forward so far. The International Linear Collider (ILC~\cite{ilc}) is a $e^+e^-$ machine based on superconducting radio-frequency cavities. The Compact Linear Collider (CLIC~\cite{clic}) has drive beam scheme capable of operating at multi-TeV energies. Both ILC and CLIC are expect to collect 100~${\rm fb}^{-1}$ after only few months of operation. A circular $e^+e^-$ collider with a circumference of approximately 80-100 km could also reach the $t \bar t$ production threshold (TLEP ~\cite{Gomez-Ceballos:2013zzn}). Research and Development towards a muon collider is also ongoing~\cite{Delahaye:2013jla}. 

The most promising method for high-precision extraction of the top quark mass is through a scan of the $t \bar{t}$ production threshold \cite{Kuhn:1980gw}. The authors of Ref.~\cite{Martinez:2002st} find that a 4-parameter fit including the top quark mass and width, the strong coupling constant and the top Yukawa coupling can yield a statistical precision of several tens of MeV on the top quark mass. Calculations of the production cross-section in the threshold region~\cite{Beneke:2008ec,Hoang:2011it,Hoang:2010gu} have since reached a precision of few percent. The potentials of ILC and CLIC have been revisited~\cite{Seidel:2013sqa} with realistic luminosity spectra for both machines, a detailed simulation of the detector response and an evaluation of the dominant systematic uncertainties. Assuming total integrated luminosity of 100~${\rm fb}^{-1}$, statistical uncertainty of 34~MeV on the (1S) top quark mass when extracted from a 10-step threshold scan was found there.  

Top quark mass measurements can also be performed at center-of-mass energies away from threshold. Above threshold (i.e. for $\sqrt{s} > 2 m_t$) the top mass extracted from the invariant mass distribution of the reconstructed top quark decay products has excellent statistical precision; Ref.~\cite{Seidel:2013sqa} quotes statistical uncertainty of 80~MeV combining the events collected in the semi-leptonic and fully hadronic decay channels for 100~${\rm fb}^{-1}$ at $\sqrt{s} =$ 500~GeV. 

The rate for single top production ($e^+e^- \rightarrow t  \bar{b} W^-$ and the charge conjugate process) depends strongly on the
top quark mass for $\sqrt{s} < 2 m_t$. The cross-section for this process is very small (less than a femtobarn for $\sqrt{s}$ below 300~GeV). Given the likely prospect that a future ILC will be operating for several years at energy around 250~GeV before any top threshold measurement can be done, an exhaustive study of the possibilities for top mass determination below threshold is highly desirable.

\subsection{Theory of $t\bar{t}$ production near threshold at $e^+e^-$ colliders}\label{sec:e+e-theory}

\begin{figure}
  \begin{center}
  \begin{tabular}{cc}
\includegraphics[width=3.2in]{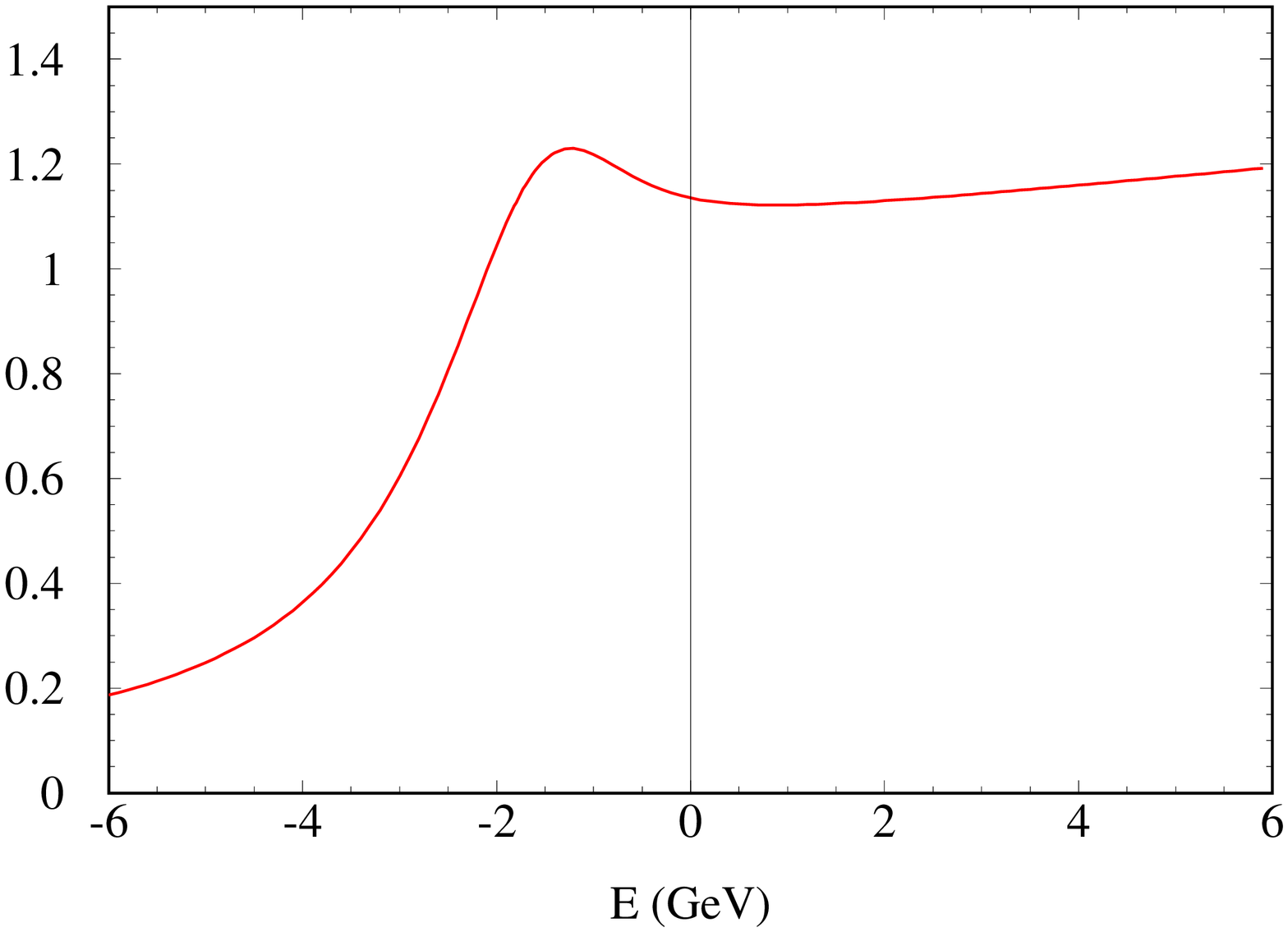} & \includegraphics[width=2.2in]{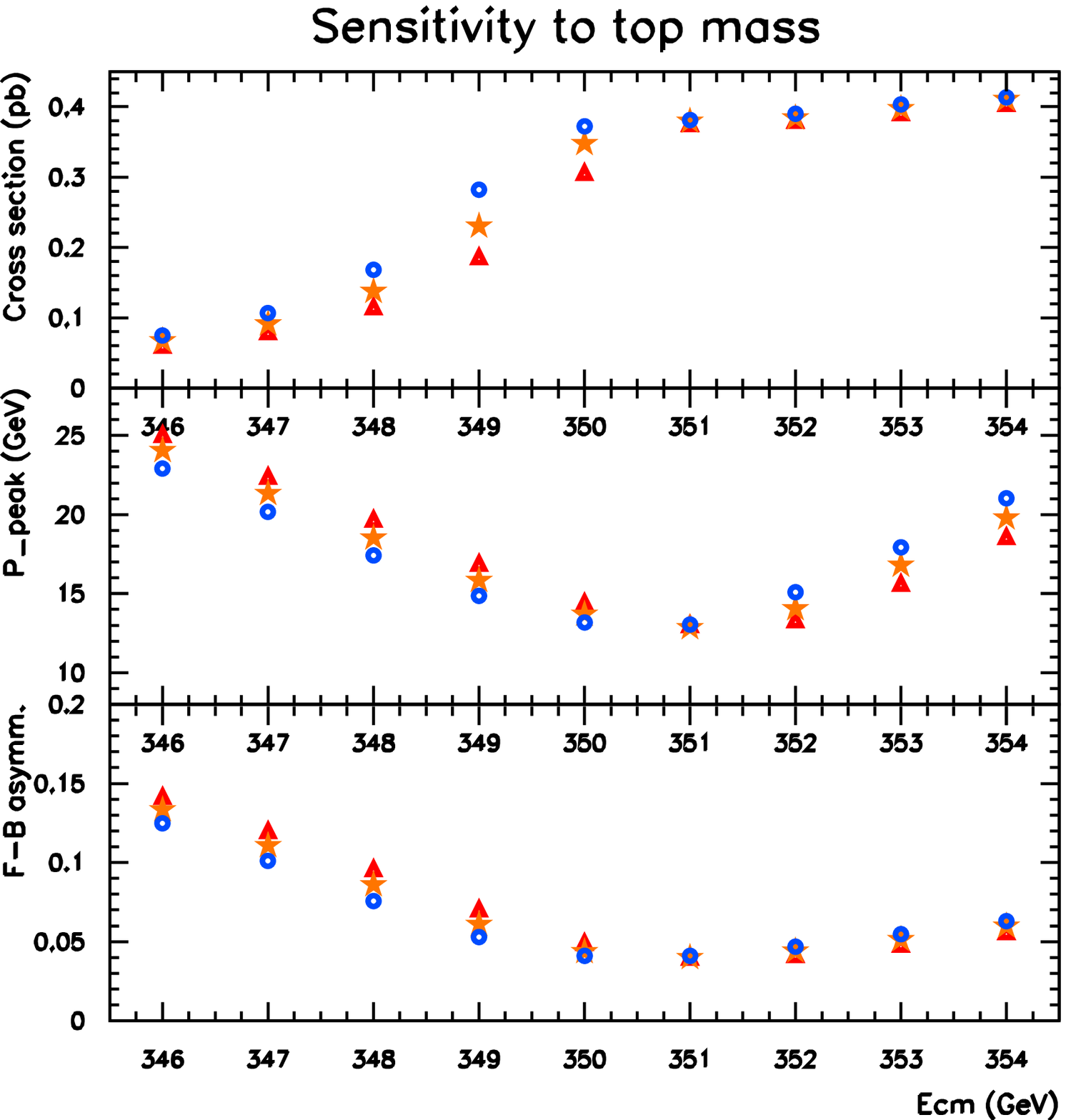} \\
 $(a)$ & $(b)$\\
\end{tabular}
  \end{center} \caption{\label{fig::crosssection} $(a)$ Typical threshold
   behavior of the (normalized) total $e^+e^- \to t\bar{t}$ cross-section with Coulomb and finite
   width effects taken into account, $E=\sqrt{s-4m_t^2}$. $(b)$ Realistic
   simulations of various observables for the $t\bar{t}$ threshold production in $e^+e^-$ collisions
   with the beam effects taken  account \cite{Martinez:2002st}. Sensitivity  to
   the top quark mass is indicated with the different symbols denoting 200~MeV
   steps in top mass.}
\end{figure}

The dynamics of top pair production at threshold is controlled mainly by two opposing effects. Firstly, due to the strong interactions, the non-relativistic quark--antiquark pair tends to form a series of Coulomb-like bound states below threshold (``toponium''). Secondly, due to the weak interactions, the large decay width of the top quark (which is comparable to its Coulomb binding energy) smears out the sharp would-be resonances in the cross-section. The interplay of these two effects leaves a single well-pronounced peak at $\sqrt {s_{res}}\approx 2m_t$ which roughly  corresponds to the would-be toponium ground state (see fig.~\ref{fig::crosssection}a). 

The expression for the resonance cross-section, $\sigma_{res}\sim\alpha_s^3/(m_t\Gamma_t)$, reveals strong dependence on the top quark mass and width as well as on the strong coupling constant. Since $\Gamma_t\gg \Lambda_{QCD}$, the top quark decays well before it hadronizes, i.e. the top quark width serves as an infrared cutoff which makes the process perturbative in the whole threshold region \cite{Fadin:1987wz,Fadin:1988fn,Strassler:1990nw}. With non-perturbative effects fully under control, perturbative QCD gives a reliable  theoretical description of the $t\bar{t}$ threshold production.

The accuracy of the approximation for $\sigma_{res}$ is limited mainly by its convergence, i.e. by the number of known terms in its perturbative expansion. Systematic calculation of the higher-order corrections in heavy quarkonium systems is based on the non-relativistic effective theory of (potential) NRQCD \cite{Bodwin:1994jh,Brambilla:1999xf,Kniehl:1999ud} which involves simultaneous expansions in the strong coupling constant and in the heavy quark velocity. The perturbative analysis has been pushed up to the NNLO by several groups \cite{Hoang:2000yr}. The NNLO corrections to the cross-section turned out to be huge despite  the renormalization group suppression of the strong coupling at the characteristic mass scales.  

A few conjectures have been made relating the slow convergence of the perturbation  theory  to the infrared renormalon contribution to the top quark pole mass, and to the  corrections enhanced by  powers of the logarithms of the heavy quark velocity in the case of the cross-section. Estimates of the missing higher order corrections have been done based on these assumptions. In particular, the use of various  ``threshold'' or ``short-distance'' mass  parameters free of infrared renormalon have been suggested in order to improve the convergence of the series for the resonance energy \cite{Hoang:2000yr}. As it turns out, however, complete control over the N$^3$LO corrections is ultimately necessary for a rigorous quantitative analysis of threshold production. Significant progress has been achieved in this field \cite{Kniehl:1999mx,Kniehl:2002br,Penin:2002zv,Kniehl:2002yv,Hoang:2003ns,Penin:2005eu,Beneke:2005hg,Marquard:2006qi,Beneke:2007gj,Beneke:2007pj,Beneke:2008cr,Anzai:2009tm,Smirnov:2009fh} and the main results are reviewed below.

\subsection{Resonance energy and top quark mass determination}

\begin{figure}
  \begin{center}
  \begin{tabular}{cc}
\includegraphics[width=3in]{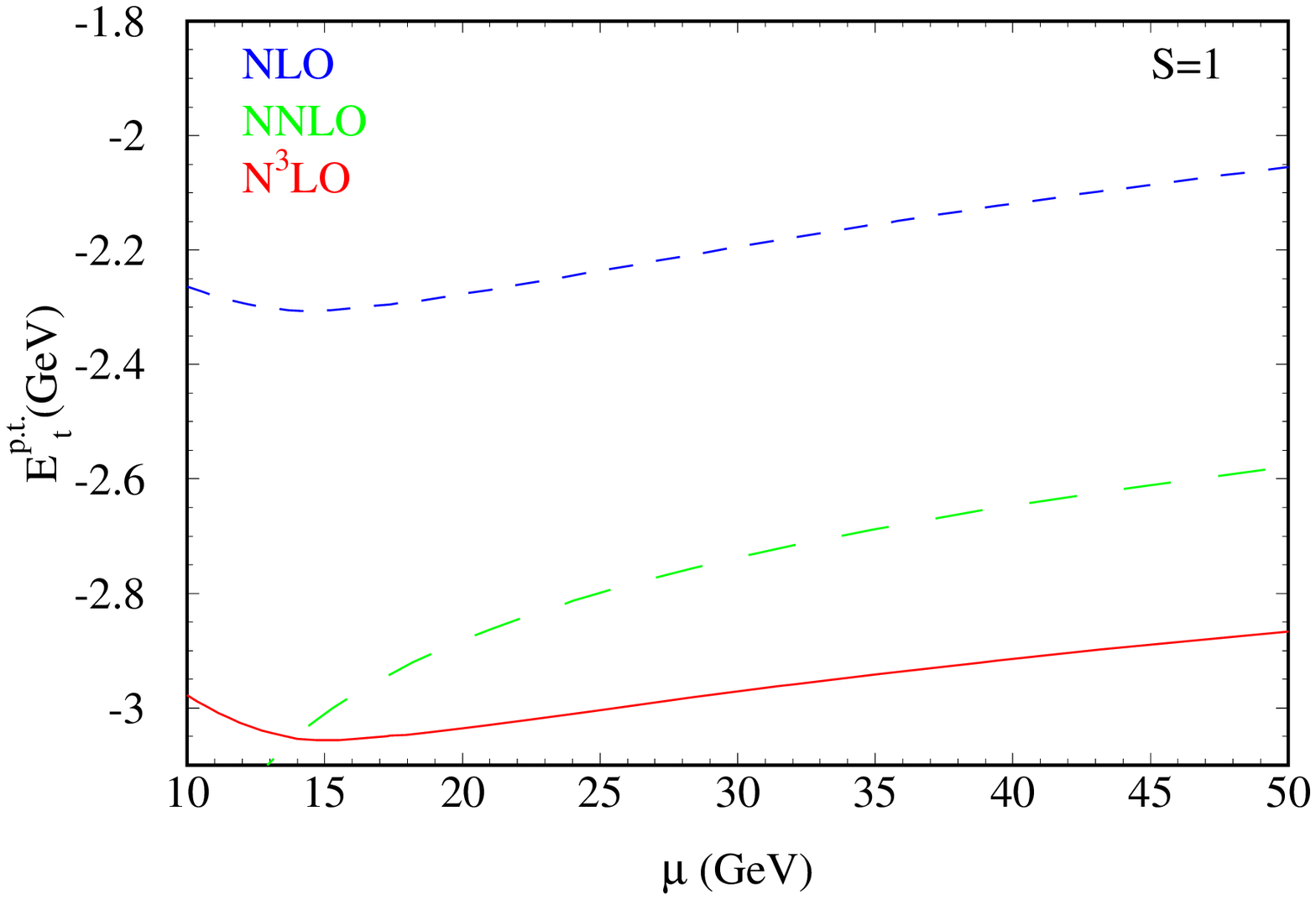}&
\includegraphics[width=3in]{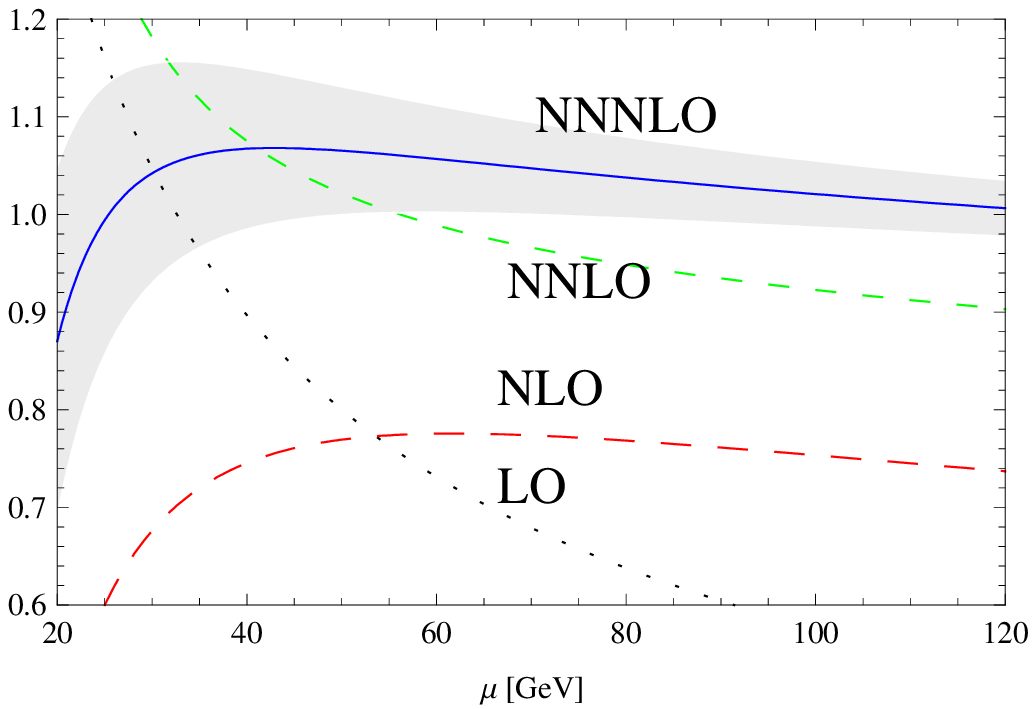}\\
$(a)$ & $(b)$\\
\end{tabular}
  \end{center}
\caption{\label{fig::convergence}  Structure of perturbative expansion for $(a)$
   the resonance energy counted from the threshold \cite{Penin:2002zv} and $(b)$
   the (normalized) resonance cross-section \cite{Beneke:2007uf} in $e^+e^- \to t\bar{t}$. Subsequent
   approximations are plotted as functions of the strong coupling normalization
   scale. The shaded area represent the uncertainty due to yet unknown
   three-loop Wilson coefficient.}
\end{figure}

The total  ${\cal O}(\alpha_s^3)$ correction to the leading order toponium ground state energy has been obtained in  \cite{Penin:2002zv}. The renormalon, logarithmic, and ``generic'' third order contributions turn out to be comparable in magnitude with no particular contribution saturating the total result.  As shown in fig.~\ref{fig::convergence}a, the third order correction stabilizes the series in the pole mass scheme and considerably reduces the scale dependence. 

The numerical analysis of Ref.~\cite{Penin:2002zv} produces a simple relation between the resonance energy and the top quark pole mass
\begin{equation}
  \sqrt{s_{\rm res}}= \left[1.9833
  + 0.007\,\frac{m_t-174.3~\mbox{GeV}}{174.3~\mbox{GeV}}
  \pm 0.0009\right]\times m_t\,,
  \label{eq::resonance}
\end{equation}
including the effect from the finite top quark width and the uncertainties in $\alpha_s(M_Z)=0.118\pm 0.003$ and from unknown high-order terms. This corresponds to a theoretical uncertainty of about $80$~MeV in the extracted pole mass. The use of a threshold mass parameter may apparently further reduce the error; for example, an uncertainty of $40$~MeV in the determination of the ``conventional'' short-distance $\overline{\rm MS}$ mass  ${\overline m}_t({\overline m}_t)$ is quoted in \cite{Kiyo:2002rr}. However, the  N$^3$LO analysis requires the ${\cal O}(\alpha_s^4)$ perturbative relation between  the pole and the $\overline{\rm MS}$ mass (currently known to ${\cal O}(\alpha_s^3)$ \cite{Chetyrkin:1999ys,Melnikov:2000qh}) and one has to rely on an assumption about the structure of the corresponding perturbative series \cite{Bauer:2011ws},  which may introduce an additional uncertainty. The calculation of the four-loop mass relation is, therefore,  crucial  for the determination of the short-distance mass with such an accuracy. At the same time the pole mass is a natural parameter for the description of the invariant mass distribution of the top quark decay product. The comparison of  the values extracted from the invariant mass distribution and from the threshold energy scan may give a realistic estimate of the experimental and theoretical uncertainties.

\subsection{Threshold cross-section}

The evaluation of the threshold cross-section through  N$^3$LO  is one of the most challenging problems of perturbative QCD. Currently the bulk of the third order corrections to the threshold cross-section is available \cite{Kniehl:1999mx,Kniehl:2002yv,Hoang:2003ns,Penin:2005eu,Beneke:2005hg,Marquard:2006qi,Beneke:2007gj,Beneke:2007pj,Beneke:2008cr} with only a few Wilson coefficients still missing. The analysis is likely to be completed in the nearest future.  

The structure of the perturbative series for the cross-section is shown in Fig.~\ref{fig::convergence}b. As in the case of the resonance energy, the third order correction stabilizes the series and the accuracy of the N$^3$LO approximation is likely to be about 3\%, or even better. Further reduction of the renormalization scale dependence may be
achieved by resummation of the higher order logarithmically enhanced corrections through effective theory renormalization group methods \cite{Hoang:2000ib,Penin:2004ay,Pineda:2006ri}. At this level of accuracy the electroweak effects become important. A  consistent treatment of the top quark finite lifetime beyond the resonance approximation has been obtained  through N$^2$LO \cite{Beneke:2010mp,Penin:2011gg}. The one-loop electroweak corrections to the cross-section have been considered in \cite{Hoang:2004tg,Hoang:2006pd}. Besides the total cross-section, differential observables including forward-backward asymmetry and the top quark momentum distribution are known through NNLO up to non-factorizable effects in the top quark finite lifetime \cite{Hoang:1999zc,Nagano:1999nw}.

\subsection{Threshold $t\bar{t}$ production at $e^+e^-$ colliders: experimental simulations}\label{sec:e+e-exp}

Realistic simulations of the $t\bar{t}$ threshold production have been performed
in \cite{Martinez:2002st}. This study assumes a 9-point energy scan around the
$t\bar{t}$ threshold where the nominal center-of-mass energy is varied between
346 GeV and 354 GeV, in 1 GeV steps, with an additional energy point taken well
below threshold to measure the background. The assumed integrated luminosity per
energy point is 30 fb$^{-1}$, for a total of 300 fb$^{-1}$ used in the full
scan. This simulation takes into account the experimental uncertainties related
to the detector effects, event selection efficiency,  and the statistics, as
well as an estimated theoretical uncertainty of 3\% in the normalization of the
cross-section. 

At each energy point, three observables are considered: the total
cross-section, the peak of the top quark momentum distribution, and the
forward-backward asymmetry. The simulations show the total cross-section to have
an estimated  experimental error of about 3\%, much below the one of the
differential observables. No theoretical uncertainties on the differential
observables have been taken into account yet. The results of the simulated scan
for these three observables are shown in fig.~\ref{fig::crosssection}b. 

As it can be appreciated, the beam energy spread, bremsstrahlung and beamstrahlung
significantly smear the measured cross-section and the precise determination of
the (machine-dependent) luminosity spectrum is crucial for the reconstruction of
the actual energy dependence of the cross-section from the threshold scan. A
multi-parameter fit including the top quark mass, top quark width and top quark
Yukawa coupling is performed considering simultaneously the three observables
mentioned above. The strong coupling constant $\alpha_s(M_Z)$ is used as an
input value with an assumed uncertainty of $\pm 0.001$. The resulting
uncertainties on the top quark mass and width are 31 MeV and 34 MeV,
respectively. Note that these estimates do not account for any uncertainties on
the nominal beam energy or the luminosity spectrum, which must be accurately
known~\cite{Djouadi:2007ik}.

More recent studies have evaluated the potential precision on the top quark mass 
considering realistic luminosity spectra generated with the {\sc GuineaPig}~\cite{GuineaPig} 
program. In particular, Ref.~\cite{Filimon} reports a detailed evaluation of the
sensitivity of the top quark mass measurement to the ILC accelerator 
parameters. The nominal ILC parameters ({\it Nominal}) are
compared to two alternative machine parameter known as {\it LowQ} and 
{\it LowP}, that have reduced and increased beamstrahlung, respectively.
Reference~\cite{Seidel:2013sqa} has compared the top quark mass extraction
form the threshold scan using luminosity spectra of the (nominal) ILC and CLIC, 
where beamstrahlung plays a more important role.

\begin{figure}[h]
\centering
\includegraphics[width=10.0cm]{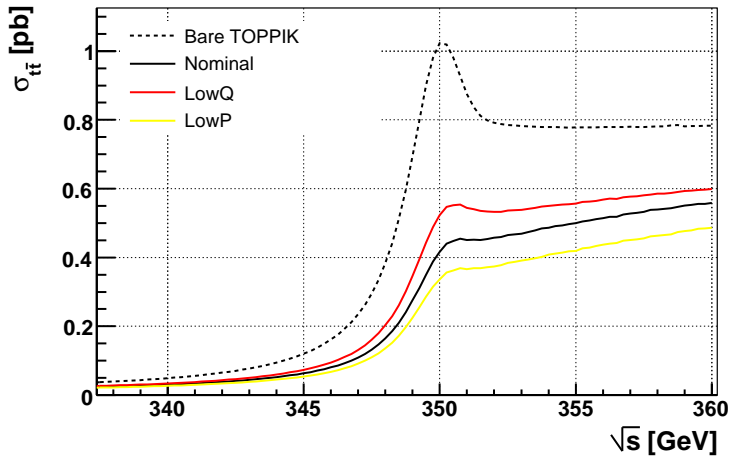}
\caption{Top quark pair production cross-section in $e^+e^-$ scattering near the $t\bar{t}$ threshold. The NNLO
   prediction based on the TOPPIK program~\cite{Hoang:1999zc}, not including
   beam effects, is shown as the dashed line. Also shown are the predicted cross
   sections after convolution of the beam effects (beam energy spread,
   bremsstrahlung and beamstrahlung) corresponding to three different sets of
   ILC accelerator parameters (see text for details).}
\label{fig:topsmeared}
\end{figure}
As an example, Figure \ref{fig:topsmeared} shows the bare $t\bar{t}$ threshold
as a function of centre of  mass energy near threshold, as well as the effective
cross-sections after convolution with the total luminosity spectrum, for the
{\it Nominal}, {\it LowQ} and {\it LowP} ILC machine parameters. The effective 
luminosity of the machine is clearly reduced due to the combined effects of 
bremsstrahlung, beamstrahlung and energy spread.
The impact on the sensitivity is rather small: the statistical uncertainty 
on the top quark mass extracted at CLIC, with a very substantial
increase in the level of beamstrahlung level, is degraded by a few MeV with 
respect to the ILC~\cite{Seidel:2013sqa}.
An accurate knowledge of the effect on the shape of the cross section
in the threshold region is however required to avoid
a large systematic contribution to the extracted mass.
While bremsstrahlung can be accurately predicted, the impact of 
beamstrahlung and beam energy spread (a much smaller contribution
to the luminosity spectrum) must be determined experimentally. 
A detailed study~\cite{Poss:2013oea} has been performed on how
to reconstruct the luminosity spectrum from Bhabha events measured
with the tracking detectors and calorimeters, taking into all relevant 
theoretical and experimental effects. This study shows that, in the context
of the CLIC accelerator at $\sqrt{s}=3$~TeV, the luminosity spectrum can be 
reconstructed to better than 5\% between the nominal and about half the nominal 
centre-of-mass energy. Pending a precise estimate of the resulting systematic 
uncertainty on the top quark mass measurement, a conservative 50 MeV uncertainty 
based on early studies is assumed here.

The uncertainty on the nominal beam energy contributes a further systematic
uncertainty. Recent studies in the context of the ILC~\cite{beamEnergy}
suggest that beam energy resolutions of $10^{-4}$ should be readily
achievable. Therefore, the uncertainty in $\sqrt{s_{res}}/2$ induced from the beam energy 
measurement is assumed to be 35 MeV and independent of luminosity and machine parameter sets. 

In summary, for a 300 fb$^{-1}$ threshold scan, the total expected uncertainty
on the top quark mass is $\sim 100$ MeV, resulting from the sum in quadrature of
the following contributions: a statistical uncertainty of order 30 MeV 
(from Ref.~\cite{Martinez:2002st}, confirmed to be possible also 
with 100 fb$^{-1}$ from a 2-parameter fit in a recent study in Ref.~\cite{Seidel:2013sqa}), 
35 MeV (beam energy), 50 MeV (luminosity spectrum) 
and 80 MeV (from the conversion of
$s_{res}$ into $m_t$ according Eq.~\ref{eq::resonance}). Given the dominance of
systematic uncertainties, it should be possible to reduce the integrated
luminosity used in the threshold scan without significantly degrading the total
uncertainty.

\subsection{Top quark mass from a reconstruction of the top decay products}

At an $e^+ e^-$ collider the top quark mass can also be measured via reconstruction in the continuum, following approaches similar to those being pursued at the Tevatron and the LHC. One could a priori hope that the cleaner environment at an $e^+e^-$ collider would  allow smaller systematic uncertainties and thus improve upon the measurements from hadron colliders.

Full simulation studies on the top quark mass via direct reconstruction at an
$e^+e^-$ collider have been carried out in both the fully hadronic ($e^+e^- \to
t\bar{t} \to q\bar{q} b q\bar{q} b$) and semi-leptonic ($e^+e^- \to t\bar{t} \to
\ell \nu b q\bar{q} b$) decay channels~\cite{ILD,SID,CLIC}. These studies have
shown that statistical uncertainties on the top quark mass below 100 MeV per
decay channel are possible assuming an integrated luminosity of 100 fb$^{-1}$ at
$\sqrt{s}=500$ GeV. A similar statistical uncertainty is obtained for the
measurement of the top width. 

Similarly to the case of hadron colliders, systematic uncertainties are again expected to be the limiting factor. At present only limited information on the anticipated experimental and theoretical systematic uncertainties at an $e^+e^-$ collider exists. Nevertheless, it is possible to obtain a rough lower limit on the total systematic uncertainty. The expected uncertainty due to fragmentation/hadronization modeling is $\sim$ 250 (400) MeV in case of the semi-leptonic
(fully hadronic) decay channel~\cite{Chekanov:2002sa}. Reconnection effects in
the final state could contribute uncertainties at the level of few hundred
MeV. Preliminary studies suggest that Bose-Einstein correlations could contribute an uncertainty of $\sim
100-250$ MeV~\cite{Chekanov:2002sa}, while color reconnection effects could also
lead to an uncertainty of $\cal{O}$(100) MeV~\cite{Khoze:1999up}. Finally, there is a theoretical uncertainty in the relation between the maximum of the invariant mass distribution and the mass parameter in the QCD Lagrangian.

It would be desirable to update these estimates taking advantage of the most recent
developments in both event generators and experimental techniques for {\it in situ}
constraining systematic uncertainties at hadron colliders. 
Taking into account all these contributions, and the fact that we have not
considered experimental systematic uncertainties ({\it e.g.} jet energy
calibration), it is difficult to imagine that the total systematic uncertainty
would be less than $(\Delta m_t)_{syst}\sim 500$ MeV, completely dominating this
measurement. Thus the threshold scan clearly beats the direct reconstruction of the top quark
mass in precision. The latter, however, can be used for additional 
control of  systematic uncertainty in the threshold measurements.

\section{Conclusions}

In the course of the 2013 Snowmass process, and during the preparation of this document, we have analyzed the theoretical and experimental aspects of the problem of top quark mass determination. We have reached the following conclusions that reflect the past developments and future prospects in this field:

\begin{itemize}
\item {\bf Need for precision in $m_t$ determination.} The current precision with which $m_t$ is known, $\delta m_t \lesssim 1\, {\rm GeV}$~\cite{TEV:2012mt,LHC:2012mt}, is already impressive; indeed the EW precision tests \cite{Baak:2012kk} are currently limited by the uncertainty in $m_W$ rather than in $m_t$. Nonetheless, motivation for increased precision may come from cosmology \cite{Bezrukov:2007ep,DeSimone:2008ei}, more fundamental issues in particle physics \cite{Degrassi:2012ry,Bezrukov:2012sa}, or a discovery of beyond the Standard Model physics at the LHC.

We estimate that some methods for top mass determination at the LHC might lead to top mass extraction with uncertainty as low as 500-600 MeV. Delivering such precision at the LHC will, however, be challenging and it remains to be seen if it can be achieved in practice. In the meantime, the most pressing issue is the relationship between the top quark mass measured at hadron colliders and a well-defined quark mass.  Meaningful improvement in the precision will therefore likely require the application of several current and novel experimental methods that are sensitive to different effects, and also advances in the theoretical understanding of the relationship between measured and fundamental quantities. 

A significant increase in precision, reaching $\delta m_t \lesssim 100\, {\rm MeV}$, can be achieved at a future lepton collider. 
\item {\bf A comprehensive collection of $m_t$ determination techniques.} This paper contains a comprehensive collection of top mass extraction methods for hadron colliders. These are methods that have been used in the past, are in current use or are under development. We discuss the salient features of each method and present estimates for the precision reach for some of them.
\item {\bf Recommendations for further studies.} Going beyond the methods discussed in this paper, we point to two problems that have not been studied so far and that we think will be playing an increasingly important role in the future. 
\begin{enumerate}
\item The possibility of BSM ``contamination" in the various top mass measurements \cite{CMPRW}. Both model--dependent and model--independent studies would be very useful. 
\item The most precise known method for extracting $m_t$ is from a threshold scan at a future lepton collider.  At present, however, it appears that the most likely lepton collider to be built is an ILC with a first stage operating at c.m. energy significantly below the $t\bar t$ threshold. The current expectation is that such first stage will be operational for a number of years; moreover, its energy upgrade might be affected by future considerations (like funding, for example). For this reason it is important to fully explore the possibility for top mass extraction at below--threshold energies through, for example, single top production. Such studies are lacking at present.
\end{enumerate}
\end{itemize}

\begin{acknowledgments}
Paper written within the Snowmass Energy Frontier working group HE3: {\it Fully Understanding the Top Quark}. We would like to thank Andre Hoang, Pedro Ruiz Femenia, Andre Sailer and Frank Simon for discussions. The work of S.~Mantry is supported by the U.S. National Science Foundation under grant NSF-PHY- 0705682. The work of A.~Mitov is supported by ERC grant 291377 ÒLHCtheory: Theoretical predictions and analyses of LHC physics: advancing the precision frontierÓ.
\end{acknowledgments}

\end{document}